\documentclass[12pt]{article}
\usepackage[margin=1in]{geometry}
\usepackage{amsmath, amssymb}
\usepackage{graphicx}
\usepackage{ccaption}
\usepackage{enumitem}
\usepackage{mleftright}
\usepackage{stackengine}
\usepackage{glossaries}
\usepackage{xcolor}
\usepackage{tikz}
\usetikzlibrary{arrows,positioning, shapes.geometric, calc}
\usepackage{subfig}
\usepackage[export]{adjustbox}
\usepackage{apacite}
\usepackage{natbib}
\usepackage{subfiles}
\usepackage{hyperref}
\usepackage{bookmark}
\usepackage{titlecaps}
\usepackage{mathrsfs}
\usepackage{fixltx2e}
\usepackage{derivative}
\usepackage{authblk}
\usepackage{float}
\usepackage{soul}
\usepackage[font={small}]{caption}

\graphicspath{ {./figs/} }

\usepackage{todonotes} 

\title{\LARGE{Learning with augmented target information: An alternative theory of Feedback Alignment}}
\author[1]{Huzi Cheng}
\author[1]{Joshua W. Brown}

\affil[1]{Indiana University, Bloomington, United States}
\date{\today}

\begin{document}

\maketitle
\begin{abstract}
    While error backpropagation (BP) has dominated the training of nearly all modern neural networks for a long time, it suffers from several biological plausibility issues such as the symmetric weight requirement and synchronous updates.
    Feedback Alignment (FA) was proposed as an alternative to BP to address those dilemmas and has been demonstrated to be effective on various tasks and network architectures.
    Despite its simplicity and effectiveness, a satisfying explanation of how FA works across different architectures is still lacking.
    Here we propose a novel, architecture-agnostic theory of how FA works through the lens of information theory:
    Instead of approximating gradients calculated by BP with the same parameter, FA learns effective representations by embedding target information into neural networks to be trained.
    We show this through the analysis of FA dynamics in idealized settings and then via a series of experiments.
    Based on the implications of this theory, we designed three variants of FA and show their comparable performance on several tasks.
    These variants also account for some phenomena and theories in neuroscience such as predictive coding and representational drift.

\end{abstract}
\section{Introduction}

For a long time, Backpropagation (BP) \citep{rumelhart1985learning, chauvin2013backpropagation} has been a dominant choice in training almost all types of neural networks and is a powerful learning algorithm.
The enormous success of BP raised ia hypothesis that the brain is doing BP in learning \citep{lillicrap2020backpropagation}.
However, through promising, this hypothesis still has a biological plausibility issue, in that to compute a synaptic change for an upstream neuron, the knowledge of all connected downstream neurons' synaptic strength is required, which is also known as the "weight symmetry problem".

Feedback alignment (FA) is an algorithm proposed to solve the problem \citep{lillicrap2014random}, which is otherwise considered to be difficult as it seems implausible to violate the locality constraint in the biological brain.
For computing the error in a given layer, FA replaces the transpose of downstream synaptic connection matrices with a fixed random matrix. With this simple modification, FA is capable of learning many tasks with different neural network structures \citep{lillicrap2014random, nokland2016direct, launay2019principled}.
The simplicity and power of this approach leads to an assumption that FA is approximating BP.
In the original study of FA \citep{lillicrap2014random}, the authors proved that in a three layer linear network, with certain assumptions, the parameter change computed with FA is aligned with the gradient computed via BP, in that the angle between BP and FA weight updates is less than 90 degrees.
Some later studies generalized these results to different extents, such as skipping layer feedback matrices, recurrent neural networks, etc., though basically all followed the same gradient-approximation framework \citep{launay2019principled, murray2019local, frenkel2021learning, refinetti2021align}.

In this paper, we revisit the question of how FA works from the perspective of information theory rather than gradient approximation.
We frame the process of FA updating parameters as augmenting the information between hidden layer neurons and targets.
First we show how FA works in an unlimited idealized hidden layer space, with the framework of information embedding. Then we examine the implications of the theory in various tasks and neural networks.
Based on the observation, we propose that FA works by augmenting the target information contained in the hidden layer, and we propose several factors that may have impacts on the performance of FA in general.
Following the proposed information theoretic perspecive on FA, we developed one trick and three variants of FA  to improve the performance of FA and push the limit of similar approaches in learning different tasks. 
Notably, one of these variants requires no information from the network output, while still being capable of training a network on classification and regression tasks. 
Lastly, we discuss these mechanisms and their potential physiological implications,  e.g., synaptic fluctuations and representational drift in the brain, and we discuss possible directions that can further improve the effectiveness of FA-related approaches.

\section{Results}

\subsection{A fixed random matrix is not approximating BP in deep nonlinear networks}

We consider a simple three-layer nonlinear neural network, $f$, with the following form:
\begin{align}
f(x) &=  W_O \cdot h(x) \\
h(x) &= \sigma (W_I \cdot x) \label{eq:w_i_function}
\end{align}

where the $x\in \mathbb{R}^d$ is the input, $h\in \mathbb{R}^m$ is the hidden layer neurons and $f(x) \in \mathbb{R}^{p}$ is the output of the whole network (sometimes also written as $\hat{y}$), $W_I \in \mathbb{R}^{m\times d}$ is the input-hidden weight, $W_O \in \mathbb{R}^{p\times m}$ is the output weight and $\sigma$ is the nonlinear activation function.
For most of our experiments, $\sigma(\cdot)$ is set to $\tanh(\cdot)$.
For simplicity, we omitted the bias term as it could be integrated into the framework by increasing the dimension of incoming neurons.

For a dataset of $n$ samples $\{x_i, y_i\}_{i=1}^{n}$ we denote them as $\mathbf{X} \in \mathbb{R}^{d\times n}$ and $\mathbf{Y} \in \mathbb{R}^{p\times n}$. The computation over the whole dataset thus is written as

\begin{align}
\hat{\mathbf{Y}} &= W_O \cdot   \mathbf{H} \\
\mathbf{H} &=\sigma ( W_I \cdot \mathbf{X} )
\end{align}

where $\mathbf{H} \in \mathbb{R}^{m \times n}$ is the hidden neuron activity across samples from 1 to $n$.

We minimize the error through Mean Squared Error (MSE) function 
\begin{equation}\label{loss_func}
  L= {\frac{1}{2}} \sum_{i=1}^{n} (\hat{\mathbf{Y}_i} - \mathbf{Y}_i)^2
\end{equation}

BP updates the gradient with $\Delta W_{O_{BP}}$ and $\Delta W_{I_{BP}}$. The former is simple to compute while the latter and can be expanded as

\begin{equation}
\Delta W_{I_{BP}} \propto \frac{\partial L}{\partial W_I} = \Delta \mathbf{H}_{BP} \cdot \sigma'(W_I \cdot \mathbf{X}_i ) \cdot \mathbf{X}_i^\top
\end{equation}

where 

\begin{equation}
\Delta \mathbf{H}_{BP} = W_O^\top (\hat{\mathbf{Y}_i} - \mathbf{Y}_i)
\end{equation}

FA solves the weight symmetry problem simply by replacing $\mathbf{H}_{BP}$ with 

\begin{equation}\label{eq:fa_signal}
\Delta \mathbf{H}_{FA} = B (\hat{\mathbf{Y}_i} - \mathbf{Y}_i)
\end{equation}

to get a corresponding $\Delta W_{I_{FA}}$ where $B \in \mathbb{R}^{m \times p}$ is independent from $W_O$ completely.

Empirically, the $\Delta W_I$ computed with $\mathbf{H}_{FA}$ works surprisingly well in reducing $L$ and it is assumed that the underlying mechanism is that 

\begin{equation}\label{w_i_angle}
\langle \mathbf{H}_{FA}, \mathbf{H}_{BP} \rangle > 0
\end{equation}

However, this only works when $\Delta W_{O_{BP}}$ is also applied to the network simultaneously.
In other words, just updating $W_I$ with $\Delta W_{I_{FA}}$ is usually not sufficient to decrease $L$. Under the assumption that FA approximates the error gradient as in BP, this lack of improvement in $L$ is not satisfying. We would rather expect Eq.\ref{w_i_angle} to hold regardless of how $W_O$ changes from the very begin if $\Delta W_{I_{FA}}$ is indeed approximating the $\partial L \over \partial W_I$.

To alternatively examine the way FA works, we adopted a binary classification task from \cite{shwartz2017opening} denoted as IBTask in this work.
In each sample of the IBTask dataset, the input $x$ is 12-d binary vector that represents a uniformly distributed point on a 2D sphere and a corresponding $y\in \{0,1 \}$.
The output $y$ is determined by a composite of a spherically symmetric real-valued function $f$ and a sigmoidal function $\psi$  $y =\psi(f(x))$ such that $P(y=1)=P(y=0)=0.5$ and the mutual information $MI(\mathbf{X}; \mathbf{Y})\approx 0.99 \text{bits}$.

Two architecturally identical networks with widths 12-8-4-2,  are trained to approximate the relationship.
$\tanh(x)$ are used as activation function for all layers except for the last one using sigmoid function.
To keep consistent with the notion we used above and the following other experiments, MSE is used as loss function.
For both of them the 12-8-4 part is treated as an ``encoder" that generates $h\in \mathbb{R}^4$ and is updated with BP, while the last layer (4-2) has errors back-propagated either through the transpose of the forward weight matrix (BP), or a fixed, random matrix (FA).  Our goal here is to test if the error signal generated by FA is informative enough to support learning.

\begin{figure}[H]
\centering
\includegraphics[width=1.0\textwidth]{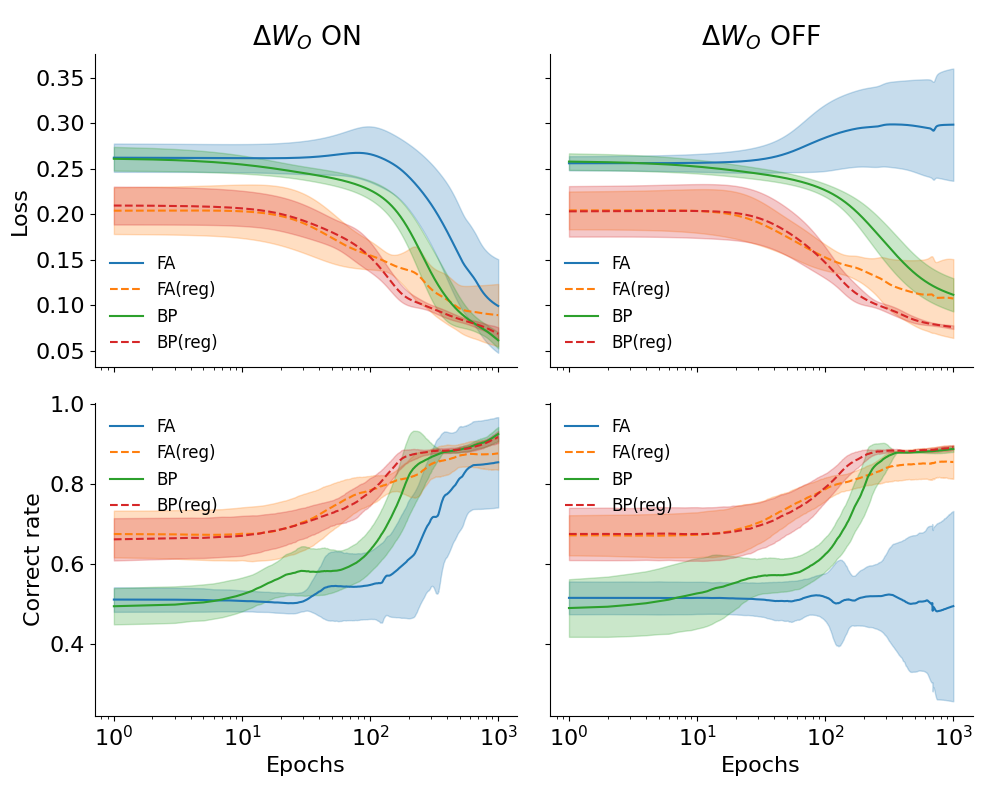}
\caption{The performance of FA and BP on the binary classification task, IBTask. Left: cases when $\Delta W_O$ is applied to networks. Right: when $\Delta W_O$ is disabled. Solid lines represent algorithm performance while dashed lines represent performance computed with linear regression.
Solid lines (FA and BP) in both sides represent the performance regarding losses and correct rates for the original networks and dashed lines (FA(reg) and BP(reg)) represent the same metrics of performance while using the output of linear regression built upon the $h$ of the original networks instead of their output layers.}
\label{fig:exp1_ibt_on_off}
\end{figure}

The results (Fig.\ref{fig:exp1_ibt_on_off}) show that, when the $\Delta W_O$ in the projections to the output layer is applied, both FA and BP reduce the loss (solid lines in upper left) to a similar level and have a correct rate of classification above 0.9 (solid lines in lower left), while FA completely fails to approximate the relationship when $\Delta W_O$ in the last layer is turned off (solid lines in upper right and lower right). This matches the observation in \cite{lillicrap2014random}, which indicates that $\Delta W_{I_{FA}}$ is not approximating $\Delta W_{I_{BP}}$ independently. 
However, we also observed that if we apply linear regression for hidden neurons' activity, $H$, in both FA and BP, their performances gets closer and increases even when $\Delta W_O$  is disabled. This result implies that FA may work independently in a way that is not approximating the gradient $\Delta W_{I_{BP}}$.

\subsection{A dynamical system idealized view of FA}\label{ds_view_FA}

The observation in Fig.\ref{fig:exp1_ibt_on_off} suggests that, though not approximating the exact gradient, parameters in the encoder, or the independent $\Delta W_{I_{FA}}$ in Eq.\ref{eq:w_i_function}, is optimizing the information between hidden layer data $H$ and output label $Y$.
To test this hypothesis, we assume that hidden states can freely move in $\mathbb{R}^{m\times n}$, i.e., the change of $H$ is independent of changes in the input data and weights, and examined how FA shapes the dynamics of $\mathbf{H}$ with the following equation in the continuous time limit (to follow the tradition of linear regression analysis, here the $\mathbf{H}$ and $\mathbf{Y}$ are the transpose of their values as used above):
\begin{equation}
\label{eq:fa_dynamics}
  \frac{d\mathbf{H}}{dt} = - B(W_O \mathbf{H} -  \mathbf{Y})
\end{equation}

As $t$ increases, it is easy to see that $\mathbf{H}$ will contain more information about $\mathbf{Y}$, which further makes performing linear regression between them easier.
We test this hypothesis by numerically simulating the first-order finite difference version of Eq.\ref{eq:fa_dynamics}.
We set $n=1000, m=6, p=1$, with all entries of $\mathbf{H}$ and $\mathbf{Y}$ independently sampled from a uniform distribution $[-0.5, 0.5]$.
$W_O, B$ are similarily initialized with $\mathcal{N}(0,1)$, while $W_O$ then is scaled by $1\over m$.

\begin{figure}[H]
\centering
\includegraphics[width=1.0\textwidth]{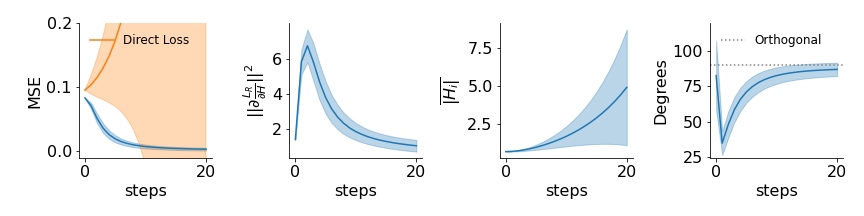}
\caption{The evolution of $\mathbf{H}$ driven by FA.
1st column: MSE calculated directly by $(W_O \mathbf{H} -  \mathbf{Y})^2$ (orange) and residual of linear regression between $\mathbf{Y}$ and $\mathbf{H}$.
2nd column: the evolution of the norm of linear regression gradient w.r.t $\mathbf{H}$.
3rd column: the evolution of the average norm of $\mathbf{H}_i \in \mathbb{R}^m$.
4th column: the evolution of the angle between the feedback signal ($\Delta \mathbf{H}_{FA}$)  and the linear regression gradient w.r.t $\mathbf{H}$ ($\frac{\partial L_{R}}{\partial\mathbf{H}}$). The horizontal grey dashed line denotes the orthogonal (90 degrees) level.
}
\label{fig:theory_fa_ds}
\end{figure}

For each step, the MSE (orange line in the first column in Fig.\ref{fig:theory_fa_ds}) 
 and the residual of the linear regression between $\mathbf{H}$ and $\mathbf{Y}$ (blue lines) are calculated as an indirect measure of information and predictability, respectively.
The results indicate that the regression residual between $\mathbf{H}$ and $\mathbf{Y}$ monotonically decreases until converging to near zero.
This implies that the feedback signal $\Delta \mathbf{H}_{FA} = - B(W_O \mathbf{H} -  \mathbf{Y})$ is approximating the gradient of linear regression given the $\mathbf{H}$. 
 We test this hypothesis by calculating the angle, i.e., the last column in Fig.\ref{fig:theory_fa_ds}, between $\Delta \mathbf{H}_{FA}$ and $\frac{\partial L_{R}}{\partial\mathbf{H}}$ where:
 
 \begin{align}
 \label{eq:grad_LR_original}
      \frac{\partial L_{R}}{\partial \mathbf{H}} &= \frac{\partial (\mathbf{H}\theta - \mathbf{Y})^\top (\mathbf{H}\theta - \mathbf{Y})}{\partial \mathbf{H}}
 \end{align}
in which $
\begin{aligned}
 \theta &= (\mathbf{H}^\top \mathbf{H})^{-1} \mathbf{H}^\top \mathbf{Y}
\end{aligned}
$ is the standard solution of the linear regression of $\mathbf{Y}$ given $\mathbf{H}$. Here $L_R$ is the loss function as in equation \ref{loss_func}, but assuming that weights $W_O$ always reflect the best least-squares linear regression between $\mathbf{H}$ and $\mathbf{Y}$.

 It is unclear how the angle between $\Delta \mathbf{H}_{FA}$ and $\frac{\partial L_{R}}{\partial\mathbf{H}}$ could be calculated analytically. Numerically, as shown in the fourth column of Fig.\ref{fig:theory_fa_ds}, the angle started around 90 degrees but quickly reduced to a much lower level and then gradually recovered to a level that is close to but constantly lower than 90 degrees.
The second phase (i.e. recovery of the angle toward 90 degrees) aligns with the convergence of regression loss $L_R$ and also the norm of $\frac{\partial L_{R}}{\partial\mathbf{H}}$ (second column in Fig.\ref{fig:theory_fa_ds}), where the alignment between the regression gradient and $\Delta \mathbf{H}$ is not necessary anymore.
Together, this indicates that the FA signal $\Delta \mathbf{H}$ is indeed optimizing the linear regression residual $L_R$ as more information of $\mathbf{Y}$ is embedded into $ \mathbf{H}$.
Therefore, one can conclude that by projecting the information of desired output label back to the hidden layer, the predictability of the output could be improved.
%
%
%


\subsection{Information augmentation in FA}

\begin{figure}[H]
\centering
\includegraphics[width=0.8\textwidth]{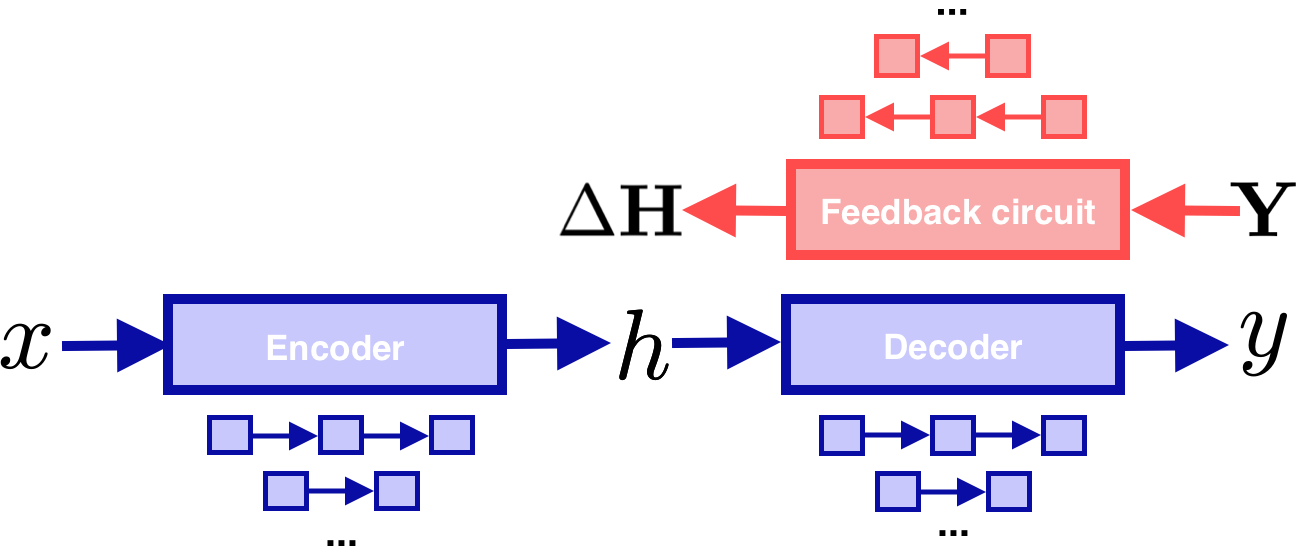}
\caption{A schematic description of how FA and its possible variants works. The encoder and decoder together form the forward path from $x$ to $y$. The feedback circuit then propagate information about $\mathbf{Y}$ to $h$ via $\Delta \mathbf{H}$. Note that for each of these components, the specific implementation is irrelevant, so they can be either a simple one layer network or deep nonlinear networks.}
\label{fig:encoder_decoder_cartoon}
\end{figure}

Based on the analysis and results above, we propose that the effectiveness of FA could be explained from perspectives other than BP approximation, because FA with $\Delta W_O$ clamped to 0 can still help learning even though the error doesn't decrease.
If one sees the linear regression in Fig.\ref{fig:exp1_ibt_on_off} as a proxy measure of the mutual information between $\mathbf{H}$ and $\mathbf{Y}$: $MI(\mathbf{H}; \mathbf{Y})$, the whole learning procedure becomes an information augmentation process:
With given $\mathbf{X}$ and $\mathbf{Y}$, all nonlinear networks that have a depth $\geq 2$ can be arbitrarily split into to two components: an encoder and a decoder in a way developed by \cite{shwartz2017opening}.
The FA then serves to gradually embed the information of $\mathbf{Y}$ into $\mathbf{H}$ by propagating error signals back, which contain information about $\mathbf{Y}$ through $e= \hat{\mathbf{Y}} - \mathbf{Y}$.
This implies that the form of $e$ may not be essential as it's just one of many possible implicit representations of the information in $\mathbf{Y}$.
More specifically, as a communication channel between $\mathbf{Y}$ into $\mathbf{H}$, the form of the feedback circuit does not matter as long as the information from $\mathbf{Y}$ is preserved.
In most cases, the dimension of the output layer $y$ is much lower than the one in the hidden layer $h$: $p \leq m$.
This makes a randomly initialized matrix, as in FA, an effective channel to send information to $h$ with a high probability, although it is not the only possible approach.
The process of transforming $\mathbf{X}$ into $\mathbf{Y}$ then becomes two independent sub-processes (see Fig.\ref{fig:encoder_decoder_cartoon}): the target of the encoder is to implement the $\Delta H$ initiated by the feedback circuit, and the objective of the decoder is to learn a better mapping based on increased mutual information $MI(\mathbf{H}; \mathbf{Y})$.
One of the simplest forms of such mapping is the linear regression used in Fig.\ref{fig:theory_fa_ds} as the linear predictability is a simple way to find how much information about $\mathbf{Y}$ is embedded in $\mathbf{H}$.
These two subprocesses can work independently as shown in Fig.\ref{fig:exp1_ibt_on_off}, where the decoder is fixed while the linear predictability is still improved over training.
Therefore, this explains why feedback alignment can work in deep nonlinear networks  \citep{nokland2016direct} and even recurrent neural networks  \citep{murray2019local} as the specific forms of the encoder and decoder are not crucial, so each of them can vary from a simple single linear mapping to a deep network.
Under this view, BP can also be treated as an encoding-decoding process. The difference between BP and other learning algorithms is that BP is a directional search, which can avoid many ``dead zone"s in $h$ that in the eyes of feedback circuits are informative but can not be realized by the encoder.
Because some choices of informative states, i.e., states that can help the decoder to reduce the final errors, in the hidden layer, could be inaccessible for the encoders no matter how their parameters are adjusted.
For example, if both $[0.8, 0.9)$ and $[1,2, 1.3)$ are informative ranges that are helpful in reducing output errors for a single neuron in the hidden layer, and if the codomain of the activation function used is $[0,1)$, then $[1.2, 1.3)$ will be a ``dead zone" for this hidden layer neuron.
In BP, since everything is calculated by partial derivatives, $[1.2, 1.3)$ will not even be set as a goal for that neuron, and thus the "dead zone" will not appear.

\subsection{Factors that contribute to the performance of FA}
\begin{figure}[H] 
    \centering
    \subfloat[]{%
        \includegraphics[width=0.3\textwidth,valign=b]{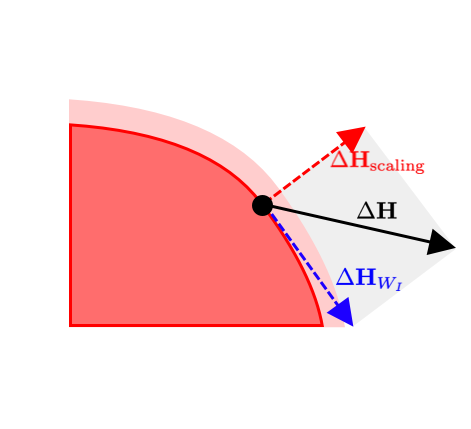}%
        \label{fig:scaling_schematic}%
        }%
    \hfill%
    \subfloat[]{%
        \includegraphics[width=0.7\textwidth,valign=b]{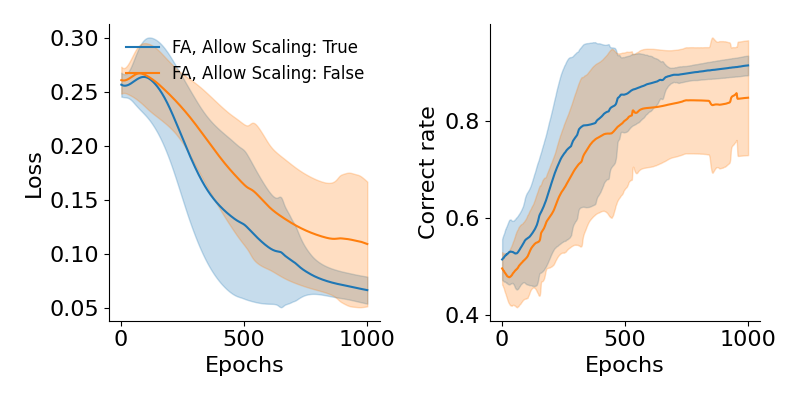}%
        \label{fig:allow_scaling_IBT}%
        }%
    \caption{The performance of FA algorithm on the Clipped polynomial approximation task with and without scaling. a: A schematic view of how $\Delta \mathbf{H}$ could be realized by both the limited activation function $\Delta \mathbf{H}_{W_I}$ and a scaling layer $\Delta \mathbf{H}_{\text{scaling}}$. b: Left: MSE as loss during training; Right: $R^2$ as performance metric during training.}
\end{figure}

Based on the above framework, we considered several factors that may contribute to the FA performance and tested them.
First, the simplified $\mathbf{H}$ evolution in Eq.\ref{eq:fa_dynamics} is only an approximation of gradient-based learning algorithms in neural networks, because the actual update equation for $\mathbf{H}$ would also have to include changes in $W_I$, or more generally changes to a series of weight parameters across potentially multiple layers before the layer $h$.
The feasible domain of this indirect update is then completely determined by the input data $\mathbf{X}$, the plausible solution space of $W_I$ and the choices of $\theta$.
Without changing $\mathbf{X}$ and $W_I$, we tested if allowing $h$ to vary beyond the hyper box determined by the $\theta$ changes the performance of FA, as activation functions are usually limited in ranges like $\tanh(x) \in (-1,1)$ or $\text{ReLU}(x) \in (0,+\infty)$.
Therefore, we added an extra scaling factor to the post-activation input.
For example, if previously the output is computed by $\text{ReLU}(W \cdot h)$, now a scaling factor $\gamma$ is inserted to the equation as  $\gamma \cdot \text{ReLU}(W \cdot h)$.
In our implementation, $\gamma$ is initialized with value 1, treated just as a freely tunable parameter and correspondingly adjusted by the same error signal using the gradient descent rule.
When the signal $\Delta \mathbf{H}$ generated by FA pushes the $\mathbf{H}$ towards the boundaries and corners of the activation function hyperbox, $\gamma$ can extend the boundary to allow more space $(-\infty, +\infty)$ to be explored by future $\Delta \mathbf{H}$ (Fig.\ref{fig:scaling_schematic}) while still preserving the nonlinearity in the network as the activation function is not changed.
This is tested with the network (Fig.\ref{fig:allow_scaling_IBT}) that is used in Fig.\ref{fig:exp1_ibt_on_off}. Using the same binary classification task, i.e., the IBTask, FA with scalable activation function outperforms the same algorithm with just normal $\tanh$.
We thus conclude that, by increasing the feasible codomain of $\mathbf{H}$ without changing the existing nonlinearity, we can expand the solution space, which then may lead to larger overlaps with informative states suggested by the FA error signal and thus makes the learning process easier.

Next, we considered the role of feedback matrices in FA.
As shown by later studies like Direct feedback alignment (DFA)\citep{nokland2016direct, launay2020direct}, several simple $B$ matrices can support learning across multiple nonlinear layers.
In DFA, the idea of random feedback signal is generalized:
For example, in a deep network with 4 hidden layers, the output error is directly transformed and passed to all hidden layers with 4 different random matrices without any chaining propagations between them, which would not be sufficient for learning with BP.
On the one hand, this makes gradient-based analysis even harder but on the other hand it suggests that instead of considering the FA as an approximation of BP, an alternative perspective from information theory might be more helpful.

As discussed above, since the information of $\mathbf{Y}$ is gradually embedded into $\mathbf{H}$, it could become easier to find a function that maps $\mathbf{H}$ back to $\mathbf{Y}$.
From this perspective, the $\Delta \mathbf{H}_{FA}$ could be viewed as an approach that improves the mutual information $MI(\mathbf{H}; \mathbf{Y})$.
Hence, any mapping, including a matrix product operation with $B$, that preserves  $MI(\mathbf{H}; \mathbf{Y})$ might be able to support learning, as long as the target information embedding process is not degenerate, i.e., the update of $\mathbf{H}$ does not collapse.
We tested this hypothesis by modifying the feedback matrix $B$ in FA with several alternative pathways:

\begin{figure}[H]
\centering
\includegraphics[width=0.7\textwidth]{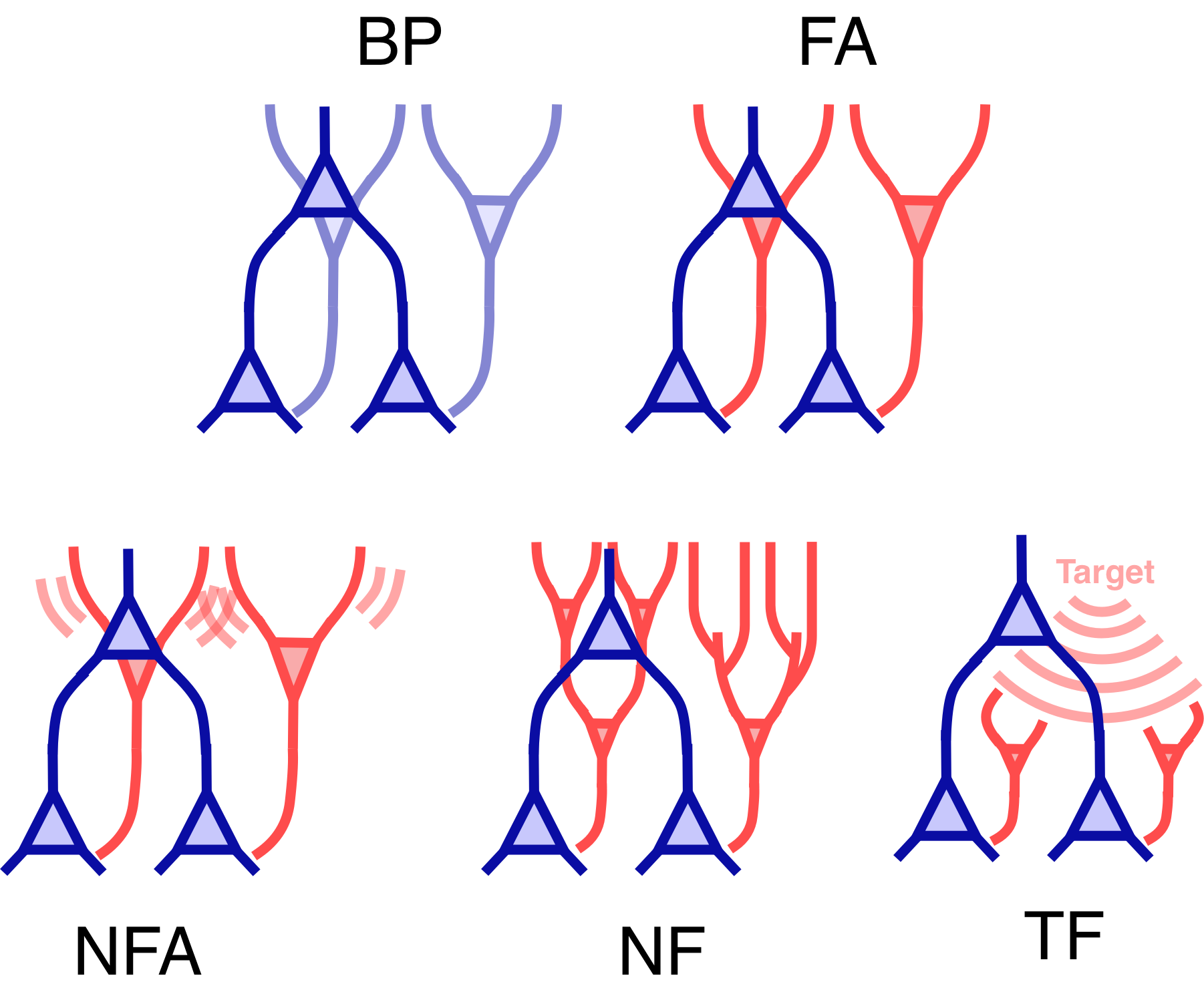}
\caption{Different types of feedback signal generation mechanisms.
First line: BP: The backward error pathway(light blue circuit) is mirroring the forward pathway (blue circuit);
FA: Using an independent backward pathway (red circuit) to propagate error signals.
Second line: \textbf{NFA}: Noisy Feedback alignment, allowing the parameters in the independent backward pathway (red circuit) to fluctuate with a gaussian noise;
\textbf{NF}: Network Feedback, using a network (the left side in the red circuit) or equivalently a multi-compartment neuron with dendrite computation (the right side in the red circuit) to propagate error signals;
\textbf{TF}: Target Feedback, dropping the error signal and only keeping the target information, propagated by a global modular signal (light red broadcasting singal) and an independent feedback pathway (red circuit).
}
\label{fig:fa_variants_cartoon}
\end{figure}
\begin{itemize}
    \item Target Feedback, \textbf{TF}: removing $\mathbf{H} \cdot W$ from the feedback signal in Eq. \ref{eq:fa_signal}. Now the error signal is just the target vector.
    This is similar to the Direct random target projection(DRTP) developed by \cite{frenkel2021learning} though it was only used with cross entropy loss function in classification tasks, where the error signal could be approximated by target signals. Considering the possibility that a feedback signal $\mathbf{Y}B$ may drive $\mathbf{H}$ to explode and the whole system then collapses and thus the entropy $I(\mathbf{H})$ drops, we augment the feedback signal to $\Delta \mathbf{H} = \widetilde{\mathbf{Y}B} + || \text{K}_{\mathbf{H} \mathbf{H}} - I||^2$, where the former $\widetilde{\mathbf{Y}B}$ denotes the $\mathbf{Y}B$ minus the average of $\mathbf{Y}B$ across each hidden neuron in $h$, playing a role that is similar to batch normalization in deep learning to keep directions evenly distributed in $\mathbb{R}^d$.
    The latter term $|| \text{K}_{\mathbf{H} \mathbf{H}} - I||^2$ pushes the correlation between different samples $\text{K}_{\mathbf{H} \mathbf{H}}$ to be nearly zero, where $\text{K}_{\mathbf{H} \mathbf{H}} =  \frac{\mathbf{H} \cdot \mathbf{H}^\top}{ |\mathbf{H}|^2} $ for each single batch of $\mathbf{H}$ .
    By introducing this term the distance between different points in $\mathbf{H}$ is increased to preserve the information $I(\mathbf{H})$ so that \textbf{TF} can effectively embed information from $\mathbf{Y}$ to $\mathbf{H}$.

    \item Network Feedback, \textbf{NF}: changing the matrix product operation to a three-layer nonlinear neural network using $\tanh$ as activation function that shares the same input-output dimension. The weights of this three-layer nonlinear network are fixed and constant, as in the linear counterpart that is FA.
    \item Noisy Feedback Alignment, \textbf{NFA}: continuously changing the $B$ matrix during the update of $\mathbf{H}$ by injecting gaussian noise into $B$, which can be viewed as re-parameterized sampling \citep{kingma2013auto} from a distribution of $B$. The injected noise does not change $B$ but instead creates a varying noisy sample of $B$.
\end{itemize}

For these three types of FA-variant mechanisms, we first tested if they can reduce the loss and improve the mutual information between $\mathbf{H}$ and $\mathbf{Y}$ by evolving $\mathbf{H}$ following Eq.\ref{eq:fa_dynamics}.
In this simulation, $\mathbf{H}$ and $\mathbf{Y}$ ($n=100000, m=6, p=1$) are independently sampled from a uniform distribution of $(-0.5, 0.5)$ are used as initial states, while other parameters are the same as ones we used in Fig.\ref{fig:theory_fa_ds}.

\begin{figure}[H]
\centering
\includegraphics[width=1.0\textwidth]{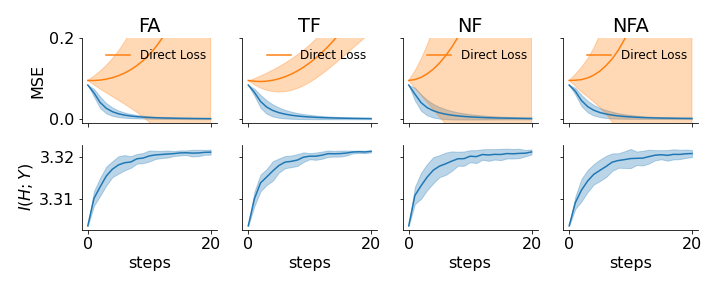}
\caption{Regression loss decreases with various types of feedback signals. \textbf{FA}: Feedback Alignment; \textbf{TF}: target feedback; \textbf{NF}: Network feedback, \textbf{NFA}: Noisy Feedback alignment.
Upper: MSE as loss function measured directly by the output error (orange) and regression loss (blue). Note that the direct loss in most cases explodes as the evolution of $\mathbf{H}$ is not bounded.  Lower: mutual information between $\mathbf{H}$ and $\mathbf{Y}$.
}
\label{fig:theory_2_MI}
\end{figure}

To calculate the mutual information, a binned method adopted from \cite{shwartz2017opening} and \cite{saxe2019information} is used and the bin size is set to 0.1 (the value of mutual information will change as the bin size changes, but the overall tendency is consistent).
The results (Fig.\ref{fig:theory_2_MI}) demonstrate that all types of feedback signals support learning regarding the regression loss, which is computed by $1\over 2n \sum_i (H_i \theta - Y_i)^2$ (the regression loss in Fig.\ref{fig:theory_2_MI}), though the direct loss using MSE exploded as the update process is not bounded.

Since the $\mathbf{H}$ here can freely move in the above simulations while the actual  $\mathbf{H}$ evolution process will be limited by the input $\mathbf{X}$, the encoder parameters and the choices of activation functions, we further examined the performance of these feedback algorithms in the above binary classification task, IBTask, from \cite{shwartz2017opening} and the MNIST dataset, MNISTTask, against the BP as a baseline.

In the binary classification task, an encoder (12-8-4) and a decoder (4-2) is used.
For the MNIST dataset, an encoder (128-64-32) built on a standard convolution backbone net (output dimension set to 128) and a decoder (32-10) is used. 
Different feedback algorithms are only applied to the last output layers (decoders) of these networks and the standard BP is applied to encoders regardless of their specific structures as our purpose is to examine the power of different feedback algorithms in improving the mutual information between $\mathbf{H}$ and $\mathbf{Y}$.

For consistency, we treat both the binary classification and MNIST dataset as regression tasks: labels are transformed into one-hot vector representations and therefore the MSE function is used as the loss function.
Among all feedback algorithms, \textbf{TF} is special since it carries no information about the performance at all, so it's natural that the weight change calculated by \textbf{TF} might be too large after convergence and lead to loss increase.
To address this issue, we use a scalar signal playing a role of global neural modulator during the learning process, to tune the magnitude of the target information.
More specifically, in these two tasks, classification accuracy is used: $\Delta \mathbf{H} = (1 - \text{accuracy})\cdot \widetilde{\mathbf{Y}B} + || \text{K}_{\mathbf{H} \mathbf{H}} - I||^2$.

\begin{figure}[H]
\centering
\includegraphics[width=0.8\textwidth]{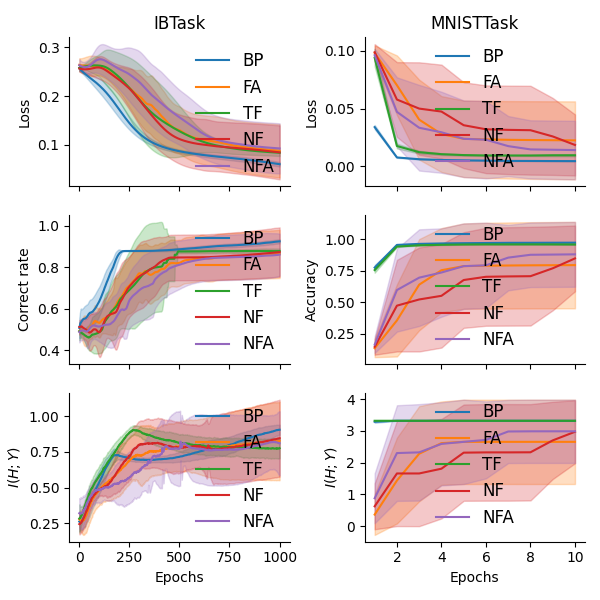}
\caption{The performance of four feedback algorithms on the binary classification (left) and MNIST(right) task.}
\label{fig:fa_variants_classification_tasks}
\end{figure}

From the results in Fig.\ref{fig:fa_variants_classification_tasks} we see that, though the feedback mechanism is drastically changed from a simple linear mapping, e.g., \textbf{NF} uses a random nonlinear mapping and \textbf{TF} drops output information, they can still reduce the loss and have comparable performance across these tasks.
Similar to \cite{shwartz2017opening}, \textbf{TF} shows indications of information compression in the IBTask (the blue line in the left side of 2nd row in Fig.\ref{fig:fa_variants_classification_tasks}), though this phenomenon has been later arguably found to be an effect of activation function choice \citep{saxe2019information}.
In our simulations, \textbf{NFA} (noisy feedback alignment) converges similarly to FA as the level of gaussian noise is set to a relatively small value thus so that across different weight updates the noise $B$ matrix can still be viewed as a stable noisy channel, with each weight element in \textbf{B} sampled as a Gaussian random variable with a fixed mean.
Besides, unlike the IBTask, in the MNISTTask, both BP and \textbf{TF} reach the peak of mutual information between $\mathbf{H}$ and $\mathbf{Y}$ as mini batch updates are used, while in the IBTask networks are trained with batch size set to the number of training data samples $n$.

IN previous work, \cite{frenkel2021learning} claim that the reason propagating solely target information back to the hidden layers works is because the target information can be viewed as a proxy of error signals in classification problems.
We extend this hypothesis by applying \textbf{TF} to regression problems.

\begin{figure}[H]
\centering
\includegraphics[width=0.8\textwidth]{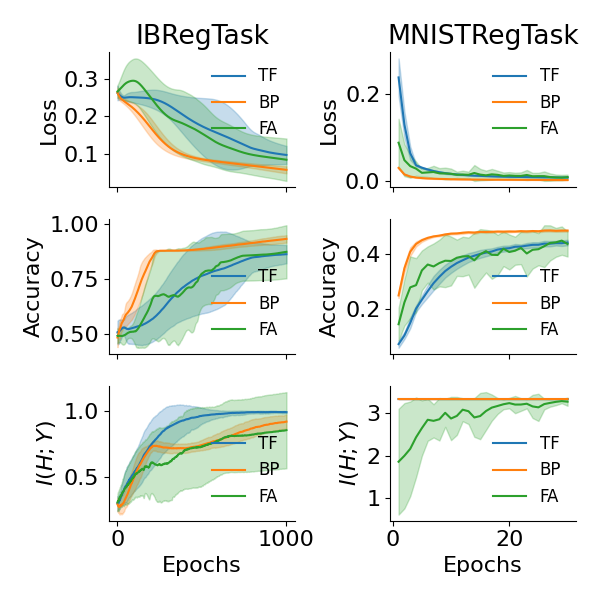}
\caption{The performance of \textbf{TF}, BP and FA on IB and MINST regression task.
}
\label{fig:tf_in_regression}
\end{figure}

The IBTask and MNISTTask are transformed into regression tasks called IBRegTask and MNISTRegTask by turning one-hot labels into a scalar between 0 and 1.
For IBTask, the output then becomes 0 and 1, while for MINSTTask the outputs then is $0.0, 0.1, \cdots, 0.9$.
To adapt to changed tasks, the decoders in these networks are respectively modified from 4-2 to 4-1 (IBRegTask) and from 32-10 to 32-1 (MNISTRegTask), while other settings are kept the same as used in Fig.\ref{fig:fa_variants_classification_tasks}.
The results (Fig.\ref{fig:tf_in_regression}) show that \textbf{TF} and FA reached similar levels of accuracy in both the IBRegTask and the MNISTRegTask.
Similar to the results in the MINST classification task (the overlapping blue and green lines in Fig.\ref{fig:fa_variants_classification_tasks}), the mutual information between $\mathbf{H}$ and $\mathbf{Y}$ in \textbf{TF} and BP quickly bumps up to the peak level after one epoch in the MNISTRegTask.

\section{Discussion}

The question of what learning algorithms are implemented in the brain has been an open question in the neuroscience community for a long time.
The speed and effectiveness of the brain in acquiring new knowledge suggests that the algorithm should be optimal. Hence BP, due to its performance supremacy, has been hypothesized to be a promising solution for learning in neuronal networks. \citep{bengio2015towards, lillicrap2019backpropagation, whittington2019theories, frenkel2021learning}.
However, though practically adopted nowadays by enormous industry-level applications and theoretically guaranteed to be sufficient in approximating functions under certain assumptions \citep{du2018gradient, allen2019convergence}, BP relies on the precise calculation of error gradients over complex functions with potentially non-local signaling, which makes it less favorable as a model for learning process in the brain.
The weight symmetry issue and the strict temporal order in calculating synaptic changes further reduces its potential in modeling how the brain learns because of the local and asynchronous nature in the brain. 

FA \citep{lillicrap2014random} sheds light on this problem by revealing that a random feedback matrix is sufficient to support propagating error signals for synaptic changes across different layers.
Later studies \citep{nokland2016direct, moskovitz2018feedback, crafton2019direct, launay2019principled} show that the idea of random feedback signals helping learning could be extended to arbitrarily distant layers in the same network and can solve large scale tasks like CIFAR  that used to be only solvable by BP.
The extreme flexibility and power shown in this family of algorithms motivated us to consider the underlying mechanism of FA and look beyond the existing studies, most of which explained FA and FA variants as approximations of BP (and specifically gradient descent) to varying extents.

The observation that the performance does not improve in a simple three-layer network trained with FA with its output weights frozen guided us to the view that the changes in the hidden layer $h$ are insufficient without the alignment of output weights.
This raises the question of whether FA creates a latent linear mapping between $\mathbf{H}$ and $\mathbf{Y}$ in the above network (the right panel in Fig.\ref{fig:exp1_ibt_on_off} $\Delta W_O$ Off case).
We observed the $R^2$ of BP and regression with $\mathbf{H}$ from FA are still increasing though $\Delta W_O$ is disabled (the solid green line and dashed orange line in the lower right panel of Fig.\ref{fig:exp1_ibt_on_off}).
This implies that the feedback signal without alignment may still be helpful for learning.
One explanation for this is that the network still works in the linear regime and thus the BP approximation theory works.
However, we didn't find a similar performance increase in the network trained with FA compared with the same network trained BP (the solid orange line in the lower right panel of Fig.\ref{fig:exp1_ibt_on_off}). 
Thus the performance increase should be explained from other perspectives other than gradient similarity.

If one sees the feedback signal propagation as imposing targets upon the hidden layer neurons, an alternative explanation naturally emerges as the process could be viewed as embedding the information of target $\mathbf{Y}$ into $\mathbf{H}$.
In the limit of loss $L$ converging to 0, the signal propagated by BP and FA will both be zero and in this case the  $\Delta \mathbf{H} \rightarrow 0$, which could be seen as hidden layer $h$ reaching targets without errors.
With this perspective, the process is converted into an information augmentation problem:
To increase the predictability between $\mathbf{X}$ and $\mathbf{Y}$, an encoder $\mathbf{X}\rightarrow \mathbf{H}$ paired with a decoder $\mathbf{H}\rightarrow \mathbf{Y}$ needs to find $\mathbf{H}$ that bridges the input and output by improving $MI(\mathbf{H}; \mathbf{Y})$ as $\mathbf{H}$ is usually determined by $x$ and does not degenerate in most cases.
Existing theories like information bottleneck theory by \cite{shwartz2017opening} and max coding rate reduction by \cite{yu2020learning} adopted this approach.
\citep{shwartz2017opening} show that BP presents such signatures of mutual information increase during training.
In this work we show that FA and its variants also show similar signatures of increased mutual information during training, and furthermore that it could be understood directly by observing the idealized $\mathbf{H}$ evolution with FA dynamics (Eq.\ref{eq:fa_dynamics}).
This further inspired us to explore whether increased mutual information effects could be generalized to account for how other learning rules might perform.  We found good generalization in this regard, as performance improved despite modifications injecting noise into $B$, using nonlinear mapping to propagate error signals and even removing errors completely from the feedback signal.

A shared characteristic of the above variants is that they can all be viewed as noisy but sufficient channels for sending information from $\mathbf{Y}$ to $\mathbf{H}$.
Their comparable performance against FA and BP across multiple tasks demonstrates the power of mutual information augmentation as a design principle in developing learning algorithms for neural networks.
Therefore, we propose that these three variants (the lower panel in Fig.\ref{fig:fa_variants_cartoon}) can be used to model a series of phenomena in the training of neuronal networks.
The \textbf{NFA} could be treated as a model that accounts for representational drift \citep{driscoll2017dynamic, attardo2018long}, in which the activation pattern of neurons changes over time while the information derived from activation patterns remains the same.
With \textbf{NFA}, the reconfiguration of $\mathbf{H}$ driven by the downstream noise could be used to exploit ``unused" space in $h\in \mathbb{R}^m$ while preserving the mutual information between the target $\mathbf{Y}$ and $\mathbf{H}$, which may facilitate long-term continuous learning.
It has also been found that a multi-compartment neuron model could be viewed as a two-layer nonlinear point-neuron network \citep{poirazi2003pyramidal, spruston2008pyramidal, losonczy2008compartmentalized}, thus in the error information pathway the circuit could be multi-layer neuronal networks, complicated neurons, or a mix of both (\textbf{NF}).
On the other hand, the encoder and decoder can also be replaced with arbitrary modules that can reduce errors, which indirectly explain why DFA\citep{nokland2016direct} works in nonlinear deep networks.
Besides, unlike highly synchronous updates in BP and the first alignment phase in FA \citep{lillicrap2014random}, the FA variants proposed in this work do not require any synchronous update as the mutual information increase $MI(\mathbf{H}; \mathbf{Y})$ contributed by the feedback pathway is independent from the decoder in \textbf{TF}.

To sum up, as an alternative to BP gradient approximation theory, we proposed an information theory based framework to explain how FA supports learning in deep nonlinear networks and even when the output/decoder parameters are frozen.
This theory furthermore allows us to predict several aspects of modifications that drastically change the structure of FA but are still effective for training. We highlighted the neurophysiological implications in these modifications.
However, some questions remain open in this framework.
First, though the theory in this work shows how FA works by mutual information augmentation and successfully predicted several variants of FA that also support learning, the analysis relies on the estimate of information in $\mathbf{H}$ and $\mathbf{Y}$, which is difficult to compute and thus can hardly be a direct objective for neural networks in realistic use cases.
Such analysis is only necessary to show how the learning laws function, and it is not necessary for the learning laws to actually function as such in a network.
It is also shown that some networks reach similar levels of performance while the corresponding mutual information is not strictly similar to each other (see the 3rd row in Fig.\ref{fig:tf_in_regression}).
To address this, a proxy becomes necessary in order to maximize the mutual information, e.g., the regression loss we used in Fig.\ref{fig:exp1_ibt_on_off} and Fig.\ref{fig:theory_2_MI}, but it is not sufficient to measure the nonlinear dependency relationships for more complicated tasks.
For more general cases, other metrics like the coding rate developed by \cite{yu2020learning} can serve similar roles.
Second, for most of tasks the dimension of input $d$ and number of neurons $m$ are much larger than the output dimension, therefore, a random mapping, like a random matrix or nonlinear random neural network, as projections from low dimensional space to high dimensional, is good enough to preserve the information of the target after mapping.
Echo state networks \citep{jaeger2007echo} for example make use of this, but we have shown previously that variant learning laws based on FA can outperform echo state networks \citep{cheng2023replay}, which suggests that large hidden layers by themselves do not optimize the information they contain simply by virtue of expanding the dimensionality of the hidden unit space.
In the future, we may explore the effect of dimension change in targets on the learning performance of different FA variants and explores ways that learn to preserve the mutual information through learning in these feedback pathways such as restricted Boltzmann machines or variational autoencoders.


\bibliographystyle{chicago}
\bibliography{ref.bib}
\end{document}